\documentclass[12pt]{iopart}

\usepackage{bm}
\usepackage{graphicx}
\usepackage{amsfonts}
\usepackage{amssymb}

\begin{document}

\title[Bose-Einstein condensation vs. localization of bosonic
quasiparticles]{Bose-Einstein condensation vs. localization of bosonic quasiparticles
in disordered weakly-coupled dimer antiferromagnets}

\author{Tommaso Roscilde$^{1,2}$, Stephan Haas$^1$ }
\address{$^1$ Department of Physics and Astronomy, University of Southern
California, Los Angeles, CA 90089-0484}
\address{$^2$ Max-Planck-Institut f\"ur Quantenoptik, Hans-Kopfermann-strasse 1,
85748 Garching, Germany}


\begin{abstract}
We investigate the field-induced insulator-to-superfluid 
transition of bosonic quasiparticles in $S=1/2$ weakly-coupled
dimer antiferromagnets. In presence of realistic disorder
due to site dilution of the magnetic lattice, we show
that the system displays an extended \emph{Bose-glass} phase
characterized by the localization of the hard-core 
quasiparticles. 
\end{abstract}
\maketitle

\section{Introduction}

 The quantum phase transition from the insulating 
to the superfluid state in strongly correlated bosonic systems
is realized in a number of model systems,
ranging from Josephson-junction arrays
\cite{FazioZ01} to cold bosons in optical lattices
\cite{Greineretal02}. Recently it has
been realized that quantum antiferromagnets also offer 
a clear example of such transition \cite{Nikunietal00}, 
and this has stimulated an intense experimental and 
theoretical activity on the subject
\cite{Rueggetal03,Rice02,Matsumotoetal04}. In the context of
quantum magnets the bosonic degrees of freedom correspond
to the local deviations from a given ground state, 
either ordered or disordered, and their condensation
corresponds to the occurrence of spontaneous 
antiferromagnetic long-range order breaking a 
planar rotational symmetry. 

 A particularly intriguing question in the context
of bosonic quantum phase transitions is their fate 
in presence of lattice disorder. General considerations
\cite{Fisheretal89} and specific examples coming from
microscopic Hamiltonians \cite{Scalettaretal91} show that a novel 
\emph{Bose-glass} phase is induced by disorder 
between the superfluid and the insulating phase,
and it corresponds to a \emph{localized} phase
for the bosonic degrees of freedom. Nonetheless,
the experimental observation of such a phase has been 
so far elusive. It appears therefore tempting to imagine
its realization in quantum magnets, where disorder
can be introduced in a highly controlled manner
through site-dilution of the magnetic lattice 
\cite{Vajketal02} and a rich variety of experimental
probes are available to detect the specific features
of this phase. 

 In this paper we show the emergence of a Bose-glass
 phase in a site-diluted $S=1/2$ Heisenberg bilayer 
 in a magnetic field. After a general description
 of the superfluid-insulator transition(s) in quantum
 magnets, we show how to recast the site-diluted
 bilayer in terms of a disordered Bose-Hubbard model
 for hard-core bosons. Then we present quantum 
 Monte Carlo data on the original spin Hamiltonian,
 clearly supporting the scenario of an extended
 Bose-glass phase in a realistic quantum magnetic model.

\section{Superfluid-insulator transitions in quantum magnets}

 Although quantum magnets are nearly perfect Mott insulators
in terms of electronic properties, they admit a
well-known bosonic representation in terms 
of local elementary spin deviations with respect 
to a given spin configuration. In the case of classically
ordered magnets, such spin deviations are the so-called 
\emph{magnons}, namely bosonic quasiparticles carrying 
a spin 1. A perfectly ordered magnet, such as a 
Heisenberg ferromagnet in its ground state, represents
therefore a trivial insulating state (the vacuum) for such 
quasiparticles. More complex ordered ground states,
in which none of the spin components reach its
saturation value, can be mapped onto 
bosonic states with a macroscopic number of bosons
displaying long-range phase coherence, namely on 
\emph{superfluid} bosonic states. The simplest
example of such a state - and the one relevant for the 
remainder of the present paper - is that of a quantum 
Heisenberg antiferromagnet in a uniform magnetic field, 
displaying canted antiferromagnetic order. 
If the field is applied along the $z$ axis, 
using, \emph{e.g.}, the Villain spin-boson transformation 
\cite{Villain74}
the deviation of the $z$ spin component from its saturation value 
at site $i$, $S-S_{z,i}$, represents
the local number of bosons $n_i$, and the trasverse spin
components $S^{\pm}_i \sim \sqrt{S} \exp(\pm j \phi_i)$
(in the limit of large spin $S$) bring the information on the 
local phase $\phi_i$. Long-range antiferromagnetic order
transverse to the field corresponds to long-range
phase coherence of the bosons \cite{Leggett95}. 
The transition from a fully ferromagnetic state
to a canted antiferromagnetic state is realized
for instance in a spin-$S$ quantum Heisenberg antiferromagnet 
on a hypercube of dimensions $D\ge 2$ when an applied uniform field 
$h = g\mu_b H / J$ (in units of the Heisenberg exchange 
coupling $J$) is driven below the critical value $h_c = 4DS$.
This transition can be conveniently described as a 
band-insulator-to-superfluid transition in the bosonic
language.

 A more complex bosonic insulating state is realized 
by quantum magnets having a quantum-disordered ground
state. This is the case of systems of $S=1/2$ antiferromagnetic
dimers with intradimer coupling $J$, coupled together
through weaker antiferromagnetic couplings $J' < J$,
with general Hamiltonian
\begin{equation}
{\cal H} = J \sum_{\langle ij \rangle} 
 {\bm S}_{i}\cdot{\bm S}_{j} + J' \sum_{\langle lm \rangle}
 {\bm S}_{l}\cdot{\bm S}_{m} - hJ \sum_{i} S^z_{i}~.
 \label{e.hamilton1}
\end{equation}
where the $\langle ij \rangle$ nearest-neighbor bonds are intra-dimer
bonds and the $\langle lm \rangle$ bonds are inter-dimer
ones. In the limit of $J'=0$ each dimer is obviously 
in a singlet state $|s\rangle = 
|\uparrow \downarrow \rangle - 
|\downarrow \uparrow \rangle)/\sqrt{2}$ with total 
spin $S_{\rm tot} = 0$.
This state naturally represents the vacuum for
three $S_{\rm tot}=1$ bosonic triplets $|t_{\pm}\rangle = 
|\uparrow \uparrow\rangle$, $|\downarrow \downarrow\rangle$, 
and $|t_{0}\rangle = |\uparrow \downarrow \rangle + 
|\downarrow \uparrow \rangle)/\sqrt{2}$, separated by a gap 
of $J$ from the singlet ground state. 
A weak coupling $J'$ between the dimers slightly perturbs 
the simple picture of dimer singlets, and the
state of each dimer is properly described as
an \emph{incoherent} mixture of the singlet state
with the triplet states, retaining the full
SU(2) symmetry of the Hamiltonian: the pure-state 
probabilities, corresponding to the diagonal 
elements of the dimer reduced density matrix 
$\rho^{\rm (dim)}$,
are $p(|s\rangle) \gg p(|t_0\rangle)= p(|t_{+}\rangle)= 
p(|t_{-}\rangle)$, and all off-diagonal terms are 
vanishing.
This picture of a so-called dimer-singlet ground state
is valid below a critical value of the ratio $g=J'/J$, 
strongly dependent on the dimensionality of the resulting 
lattice and on the specific geometry of the weak couplings.
In what follows we assume the system to be well inside
the dimer-singlet regime in zero field.

Applying a magnetic 
field to each isolated dimer ($J'=0$) along the quantization 
axis, the two triplets $|t^{-} \rangle$ and 
$|t^{0} \rangle$ remain well separated from the
singlet ground state, while the $|t^{+} \rangle$
triplet is brought to degeneracy with the singlet 
for a field $h=1$, at which the system jumps
to a fully magnetized state. In presence of a finite 
interdimer coupling $J'>0$ the application of a weak  
field $h \lesssim 1$ reduces the symmetry of the Hamiltonian 
from SU(2) to U(1), and correspondingly in the dimer-singlet 
phase the reduced density matrix $\rho^{\rm {(dim)}}$ for each dimer 
has $p(|t_0\rangle) \neq  p(|t_{+}\rangle)= 
p(|t_{-}\rangle)$, but the last equality  
guarantees that the system does not develop a finite
magnetization along the field, so that the 
ground state has a higher symmetry than the
Hamiltonian. Moreover the mixture remains
incoherent to mantain the U(1) symmetry. 

Nonetheless, the interdimer $J'$ interaction 
is clearly seen to couple effectively the singlet
state with the $t_{+,-}$ triplets through terms
of the type $J' S_l^{\pm} S_m^{\mp}$ where $l$ is one
of the two sites of the dimer and $m$ is a site
of a neighboring dimer (actually the $J'$ term
couples coherently the $t_{+,-}$ triplets and 
the singlet on both neighboring dimers). When the singlet
and the $t_+$ triplet are brought close to
degeneracy by the field, this coupling becomes resonant 
and it changes drastically the property of 
the ground state. A critical field 
$h^{(0)}_{c1} \sim  J- aJ'$ (where $a$ is a model-dependent 
constant) suffices to match the
resonance conditions. The resonant coupling between
the two states causes the breaking of all
the symmetries previously retained by the reduced 
density matrix. A transfer of 
population from the singlet to the $t_+$ triplet
breaks the symmetry between $t_+$ and $t_-$,
thereby leading to the appearence of a finite
magnetization along the field. Moreover the
reduced density matrix acquires finite
off-diagonal terms  
$\langle s | \rho^{\rm {(dim)}} | t_+ \rangle = 
\langle t_+ | \rho^{\rm {(dim)}} | s \rangle \neq 0$,
which implies that, defining creation/annihilation operators
for the $t_+$ bosons $b^{\dagger}_{t_+}|s\rangle = | t_+ \rangle$,
$b_{t_+}|t_+ \rangle = | s \rangle$, the expectation 
values $\langle b^{\dagger}_{t_+} \rangle$ and
$\langle b_{t_+} \rangle$ become non-zero. This clearly
identifies the breaking of the U(1) symmetry in the 
plane (\emph{i.e.} the appearence of antiferromagnetic 
order transverse to the field) with the appearence of 
a condensate of $t_{+}$ bosons. 

 From the magnetic point of view, the field-induced
 ordered ground state is a canted antiferromagnetic
 state of the type described at the beginning of
 the section, with finite uniform magnetization along 
 the field and finite transverse staggered magnetization.
 By further increasing the field, the system is eventually
 brought to a fully polarized ferromagnetic state
 when reaching a critical field $h^{(0)}_{c2} \sim J + \bar{a} J'$
 (with $\bar{a}$ again model-dependent) which 
 destroys the coherent mixing between the singlet and
 the $t_+$ triplet. Around this field the dimer reduced density
 matrix has $p(|t_+\rangle) \gg p(|s\rangle)$, so the 
 ordered state is conveniently described as a superfluid 
 state of bosonic singlet holes in the triplet "sea".
 The fully polarized state corresponds simply to the
 vacuum of such holes. Therefore we can conclude that
 a weakly coupled 
 dimer system in a field realizes two successive 
 insulator/superfluid transitions.

\begin{figure}[h]
\begin{center}
\includegraphics[
     width=160mm,angle=0]{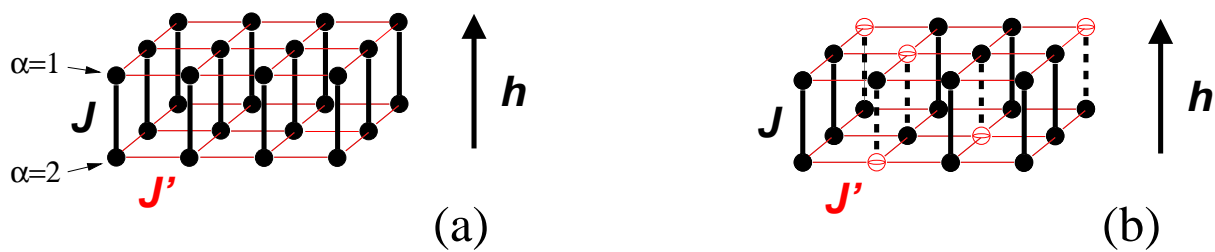} 
\caption{(a) Heisenberg antiferromagnet on a bilayer
and in a magnetic field; $\alpha$ represents the layer index.
(b) Same as in (a) but with site dilution. The open
dots correspond to missing sites, and dashed
$J$ bonds correspond to dimers that have lost 
one spin.}
\label{f.structure}
\end{center}
\end{figure}

\section{Bilayer Heisenberg antiferromagnet in a field}

 In what follows we consider the specific case of 
 a $S=1/2$ Heisenberg bilayer in a uniform magnetic field
 [Fig. \ref{f.structure}(a)]:
\begin{equation}
{\cal H} = J' \sum_{\langle ij\rangle} \sum_{\alpha=1, 2}
 {\bm S}_{i,\alpha}\cdot{\bm S}_{j,\alpha}  
+ J \sum_{i}  {\bm S}_{i,1}\cdot{\bm S}_{i,2}
 - Jh \sum_{i,\alpha} \epsilon_{i,\alpha} S^z_{i,\alpha}~.
 \label{e.hamilton2}
\end{equation}
where the index $i$ runs over the sites of a square lattice, 
$\langle ij\rangle$ are pairs of nearest neighbors on the 
square lattice, and $\alpha$ is the layer index.
This model has been intensively investigated
theoretically \cite{SandvikS94,Sommeretal01} and it also
captures the main features of the magnetic interactions 
in the recently synthetized compound BaCuSi$_2$O$_6$
\cite{Jaimeetal04}.

 From the qualitative discussion of the previous section
we have seen that the insulator/superfluid transition
in a weakly-coupled dimer system involves essentially
the singlet and one of the triplet states ($t_+$) of each 
dimer. Within a low-energy effective description,
appropriate close to the degeneracy point between
these two states, it is then reasonable to discard the 
other two triplet states ($t_+$) and derive an effective
dynamics for the triplet $t_+$ bosons only (or, equivalently, 
for the singlet bosonic holes). To retain the ground
state symmetries of the dimer-singlet phase after discarding 
two of the triplets,
it is necessary to assume that the ground state below 
the lower critical field $h^{(0)}_{c1}$ is a product of 
singlets $|\Psi\rangle = \prod_i |s_i\rangle$, so that the 
bosonic insulating state is simply the vacuum for the 
$t_+$ triplets. 

 The derivation of the effective bosonic Hamiltonian
is well known \cite{TachikiY70} and here we simply sketch it. 
Introducing the operators 
${\bm s}_i = {\bm S}_{i,1} + {\bm S}_{i,2}$ and 
${\bm t}_i = {\bm S}_{i,1} - {\bm S}_{i,2}$ the 
intradimer Hamiltonian reads 
${\cal H}_{\rm intra} = (J/2) \sum_i ({\bm s}_i^2 - 
h s_i^z)$ (neglecting an irrelevant additive constant) 
and the interdimer one reads 
${\cal H}_{\rm inter} = (J'/2) \sum_{\langle ij\rangle}
({\bm s}_i \cdot {\bm s}_j +{\bm t}_i \cdot {\bm t}_j)$.  
Truncating the dimer spectrum to $|t_+\rangle$
and $|s\rangle$, one introduces the pseudospin 
states $|\uparrow\rangle = |t_+\rangle$,
$|\downarrow\rangle = |s\rangle$, and associated
$S=1/2$ pseudospin operators $\sigma_i^{z}$, $\sigma_i^{\pm}$.
The ${\bm s}_i$, ${\bm t}_i$ operators are then
expressed in terms of the pseudospin ones as:
$s_i^z = \sigma_i^{z} + 1/2$, $t_i^{\pm} = \sqrt{2}
\sigma_i^{\pm}$, and all other components are zero
within the subspace spanned by $|s\rangle$ and $|t_+\rangle$.
By further exactly mapping the pseudospins
onto hardcore bosons, $\sigma_i^z = n_i - 1/2$ 
($n = b^{\dagger}b$), 
$\sigma^{+}_i = b_i^{\dagger}$, and $\sigma^{-}_i = b_i$,
with $[b_i,b^{\dagger}_j] = 0$ ($i\neq j$) and
$\{b_i,b^{\dagger}_i\} = 0$,
we obtain the effective hard-core bosonic Hamiltonian
\begin{equation}
{\cal H}_{\rm boson} = 
-J'/2 \sum_{\langle ij \rangle} (b_i b_j^{\dagger} + h.c.)
+ J'/2 \sum_{\langle ij \rangle} n_i n_j - J(h-1) \sum_i n_i
\end{equation}
On each dimer location $i$, $|0\rangle = |s\rangle$ and
$|1\rangle = |t_+\rangle$. For $h < h^{(0)}_{c1} = J - zJ'/2$
the ground state is the vacuum, while for 
$h > h^{(0)}_{c2} = J + zJ'/2$ the ground state has one
hardcore triplet per dimer ($z=4$ is the coordination
number of the square lattice). Both insulating states
have a \emph{finite particle gap}. For any intermediate 
field value $h^{(0)}_{c1} \leq h \leq h^{(0)}_{c2}$
the system is \emph{gapless} with bosons having a 
fractional filling (which corresponds to a finite 
uniform magnetization away from its saturation value,
$0 < m_u^z = \langle S_i^z \rangle < 1/2$), and 
forming a condensate with long-range phase coherence 
(which corresponds to 
staggered antiferromagnetic order transverse to
the field $m_s^{x(y)} = \langle (-1)^i S_i^{x(y)}\rangle$).

\begin{figure}[h]
\begin{center}
\includegraphics[
     width=80mm,angle=0]{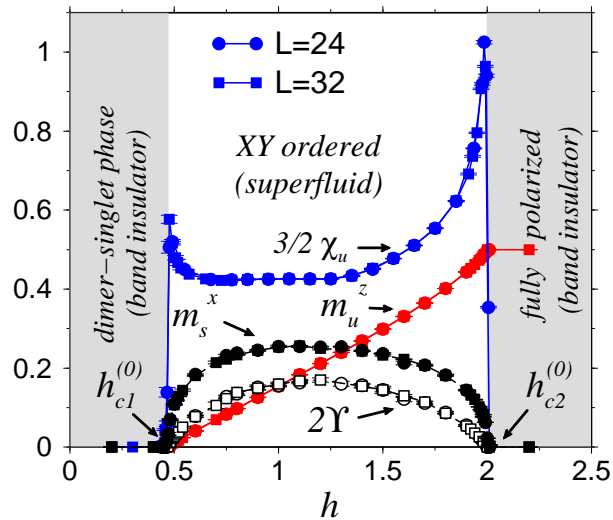} 
\caption{Field dependence of the uniform magnetization
$m_u^{z}$, uniform susceptibility $\chi_u$, 
transverse staggered magnetization $m_s^{x}$,
and superfluid density $\Upsilon$ at $T=0$ for the 
bilayer antiferromagnet with $J/J'=4$ in the clean limit.}
\label{f.clean}
\end{center}
\end{figure}

 Fig. \ref{f.clean} shows the succession of ground-state 
bosonic phases in a bilayer Heisenberg antiferromagnet with 
$g = J/J' = 4$. Shown are the results of a quantum Monte Carlo
calculation based on the Stochastic Series Expansion 
method with directed-loop update \cite{SyljuasenS02}.
The superfluid density $\Upsilon$ is estimated through winding
number fluctuations \cite{PollockC87}, while the
uniform susceptibility $\chi_u$ is obtained by
numerical derivation of the uniform magnetization
with respect to the field.

\section{Bilayer Heisenberg antiferromagnet with site dilution: Bose-glass phase}

A well controlled way of introducing lattice disorder
in quantum spin systems is by doping the magnetic ions
with non-magnetic ones with the same valence, so that the
main effect is the removal of some of the spins in the
magnetic Hamiltonian. In the case of BaCuSi$_2$O$_6$, 
\emph{e.g.}, this can be achieved by doping the Cu$^{2+}$
ions with non-magnetic ions as Zn$^{2+}$ or Mg$^{2+}$.
The magnetic Hamiltonian of the Heisenberg bilayer 
in presence of site dilution  [Fig. \ref{f.structure}(b)] reads: 
\begin{equation}
{\cal H} = J \sum_{i} \epsilon_{i,1} \epsilon_{j,2} 
 {\bm S}_{i,1}\cdot{\bm S}_{i,2} 
 + J' \sum_{\langle ij\rangle} \sum_{\alpha=1, 2}
\epsilon_{i,\alpha} \epsilon_{j,\alpha} 
 {\bm S}_{i,\alpha}\cdot{\bm S}_{j,\alpha}  
 - Jh \sum_{i,\alpha} \epsilon_{i,\alpha} S^z_{i,\alpha}~.
 \label{e.hamilton3}
\end{equation}
Here the variables $\epsilon_{i,\alpha}$ are random
numbers taking value 0 with probability $p$ (corresponding
to the concentration of non-magnetic dopants) and 1
with probability $1-p$.

 In presence of site dilution, some of the dimer are
 completely missing, and some are reduced to single 
 spins. We can nonetheless repeat the same approximate
 bosonic mapping as discussed in the previous section
 by truncating the Hilbert space of the intact dimers
 while retaining the full space of the dangling spins.
 The $J'$ coupling between two intact dimers has obviously
 the same expression in terms of bosonic operators as before.
 The coupling between a dimer and a dangling spin on the
 $\alpha$-th layer is instead of the form 
 ${\cal H}_{\rm dimer-spin} = (J'/2)~({\bm t}_i + {\bm s}_i)\cdot {\bm S}_{j,\alpha}$.
 Introducing the pseudospin operators for the dimers, 
 we obtain 
${\cal H}_{\rm dimer-spin} = 
(J'/\sqrt{2}) ( \sigma_i^{x} S_{j,\alpha}^x + 
\sigma_i^{y} S_{j,\alpha}^y) + (J'/2)(\sigma_i^{z}+1/2)S_{j,\alpha}^z$.
Finally, mapping the
pseudospins onto $b, b^{\dagger}$ 
hardcore bosons and the dangling $S=1/2$ spins
onto $c, c^{\dagger}$ hardcore bosons, we end up with the following
(rather complex) disordered bosonic model
\begin{equation}
{\cal H}_{\rm boson} = {\cal H}_{bb} + {\cal H}_{cc} + {\cal H}_{bc}
\end{equation}
where
\begin{eqnarray}
\null 
{\cal H}_{bb} &=& 
-\frac{J'}{2} \sum_{\langle ij \rangle} \lambda_i \lambda_j~ (b_i b_j^{\dagger} + h.c.)
+ \frac{J'}{2} \sum_{\langle ij \rangle} \lambda_i \lambda_j~  n_i n_j - 
J(h-1) \sum_i \lambda_i n_i ~~~~~\\
{\cal H}_{cc}
&=& - \frac{J'}{2} \sum_{\langle ij \rangle, \alpha}
\gamma_{i,\alpha} \gamma_{j,\alpha}~ (c_i c_j^{\dagger} + h.c.)
+ J'  \sum_{\langle ij \rangle} 
\gamma_{i,\alpha} \gamma_{j,\alpha}~ M_i M_j  \nonumber \\
&& -  
\left[Jh + \frac{J'}{2} \sum_d (\gamma_{i+d} - \lambda_{i+d})\right]
\sum_i \gamma_i ~M_i  \\
{\cal H}_{bc} &=& -\frac{J'}{2\sqrt{2}} \sum_{\langle ij \rangle}
\left[ \lambda_i \gamma_j ~(b_i c_j^{\dagger} + h.c.)
+ \gamma_i \lambda_j ~(c_i b_j^{\dagger} + h.c.) \right] \nonumber \\
&& +  J'  \sum_{\langle ij \rangle} 
(\lambda_i \gamma_j~ n_i M_j + \gamma_i \lambda_j~ M_i n_j).
\end{eqnarray}
Here $M_i = c_i^{\dagger} c_i$, 
$\gamma_{i,\alpha} = \epsilon_{i,\alpha}(1-\epsilon_{i,\bar{\alpha}})$,
$\gamma_i = \sum_{\alpha} \gamma_{i,\alpha}$, and 
$\lambda_i = \epsilon_{i,1} \epsilon_{i,2}$ and $\sum_d$ 
runs over the four lattice vectors. 

 For site dilution well below the bilayer
 percolation threshold ($p<p^{*} = 0.5244(2)$ \cite{RoscildeH05}), 
 the above model contains a network of $b$-sites 
 ($\lambda_i = 1$, $\gamma_i = 0$),
 corresponding to intact dimers in the original spin Hamiltonian,
 interrupted by $c$-sites 
 ($\lambda_i = 0$, $\gamma_i = 1$), corresponding to dangling 
 spins in the original model, and by empty sites
 ($\lambda_i = 0$, $\gamma_i = 0$) corresponding to
 missing dimers. 
 The most relevant feature of this complex model is that 
 the chemical potential for the $b$-bosons,  
 $\mu_b =  J(h-1)$ is quite different
 from that of the $c$-bosons, 
 $\mu_c = Jh + \frac{J'}{2} \sum_d (\gamma_{i+d} - 
 \lambda_{i+d})$. In fact, if $J \gg J'$,  
 the chemical potential for the $c$-bosons is
 always positive, and when $hJ \sim J'$ the 
 $c$-sites are almost all filled, which
 corresponds to field polarization of the dangling 
 spins. 
 On the contrary the chemical potential
 for the $b$-bosons can be negative for $h<1$, 
 which means that, for the same field $h$, a system 
 with only $b$-sites can be empty of 
 bosons, while a system with only $c$-sites would 
 be completely filled. When $b$- and $c$-sites 
 are combined together, $c$-bosons can hop to neighboring 
 $b$-sites, so that there is always a finite, albeit small, 
 population
 of $b$-bosons around $c$-sites, but, for negative enough $\mu_b$,
 bosons are prevented from traveling deep inside 
 continuous regions of $b$-sites. Therefore
 for $h \lesssim h^{(0)}_{c1}$ the bosonic model has
 all $c$-sites essentially filled and $b$-sites 
 only partially filled when neighboring a $c$-site,
 but continuous regions of $b$-sites are essentially
 empty of particles. If $J\gg J'$, in this situation 
 the system has a gap to the addition of further 
 particles, and it is in a quantum disordered state very
 similar to the insulating states in the clean limit.
 
 When $h \geq h^{(0)}_{c1}$ the first $b$-bosons appear
 in a homogeneous system made exclusively of $b$-sites,
 and they form a coherent condensate. In presence of
 disorder, instead, $b$-bosons are induced away from 
 $c$-sites only in those rare regions which are 
 locally well approximating a clean system, namely where
 the local coordination number is very close to $z=4$
 of the clean system. In regions with a lower local 
 coordination number the local critical field is
 effectively higher, given its dependence on $z$.
 This simply means that the first bosons to appear in
 the bulk of the $b$ regions are localized, although
 the $b$-sites might form a percolating cluster for
 low enough dilution of the lattice.
 Therefore we are in presence of a phenomenon of 
 \emph{quantum localization} of the $b$-bosons, introducing
 a new phase in the disordered system, namely a 
 \emph{Bose-glass} phase. In such a phase there is
 no gap to the addition of a particle because, for 
 any value of $h \gtrsim h^{(0)}_{c1}$, we will find
 a region in the system where the local critical field
 equals $h$, so we can inject a boson there. 
 This is therefore an unconventional 
 quantum-disordered insulating phase
 with short-range correlations and a gapless spectrum. 
 To overcome localization and enter the condensate phase, 
it is necessary to reach a higher critical field 
$h_{c1} > h_{c1}^{(0)}$ at which a percolating network
of $b$-sites becomes accessible to bosons, so that 
long-range phase coherence can be established throughout
the system. 
 
  A similar phenomenological description can be 
 repeated when approaching the upper critical field, 
 $h \sim h^{(0)}_{c2}$. Here the description is simplified
 by the fact that, for such a high field, the $c$-sites
 can be thought of as perfectly filled, and bosonic
 holes are all on $b$-sites (a bosonic hole on a 
 $c$-site is energetically quite unfavorable).
 Therefore the $c$-sites can be simply quenched
 in the $|M_i = 1\rangle$ state, and we get the simplified bosonic
 model
\begin{equation} \null\hspace{-2.3cm}
{\cal H}_{\rm boson} \approx
-\frac{J'}{2} \sum_{\langle ij \rangle} \lambda_i \lambda_j~ (b_i b_j^{\dagger} + h.c.)
+ \frac{J'}{2} \sum_{\langle ij \rangle} \lambda_i \lambda_j~  n_i n_j - 
\sum_i \lambda_i \left[ J(h-1)  - J' \sum_d \gamma_{i+d} \right] n_i~~
\end{equation}
which is the hardcore Bose-Hubbard model on a site-diluted
square lattice with random on-site chemical potential
(fluctuations in the chemical potential appear
only close to a site vacancy). 
As before, the regions of $b$-sites with a lower local 
coordination number will have a local upper critical field which
is less than $h^{(0)}_{c2}$, which means that they tend
to expel bosonic holes. For 
$h \lesssim h_{c2}^{(0)}$ such holes
remain localized on regions with local coordination close
to that of the clean system, implying that, for a critical
field $h_{c2} < h_{c2}^{(0)}$, the system loses the hole
condensate and it enters a Bose-glass phase
of bosonic holes.

\section{Diluted Heisenberg bilayer: quantum Monte Carlo data}

 The approximate bosonic picture we put forward 
in the previous section for the physics of the site-diluted
Heisenberg bilayer is confirmed by extensive quantum
Monte Carlo simulations done on the original spin model
making. Making use of Stochastic-Series-Expansion quantum 
Monte Carlo, we were able to efficiently reach the physical 
$T=0$ behavior by a successive increase of the inverse
temperature \cite{Sandvik02}. We investigated system
sizes up to $40\times 40\times 2$, averaging typically
over 200 disorder realizations. 

\begin{figure}[h!]
\begin{center}
\includegraphics[
     width=100mm,angle=0]{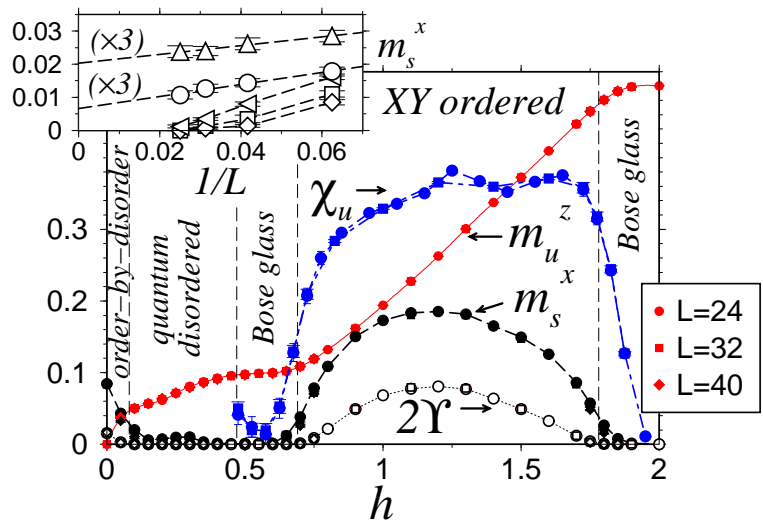} 
\caption{Zero-temperature field scan in 
the site-diluted bilayer Heisenberg antiferromagnet 
with $J/J'=4$ and $p=0.2$. Inset: scaling of the order parameter
$m_s^x$ in the different phases of the system; from top
to bottom $h=0.75$ (XY ordered), $0.05$ (order-by-disorder), $1.85$
(high-field Bose glass), $0.35$ (quantum disordered), $0.6$
(low-field Bose glass). }
\label{f.hscan}
\end{center}
\end{figure} 

 Fig. \ref{f.hscan} shows the succession of phases
in a bilayer Heisenberg model  with $J/J'=4$ and 
site dilution $p=0.2$
when the applied field $h$ is changed from zero 
to the saturation value $h=2$. This picture has to 
be contrasted with the analogous field scan in the 
clean limit, as shown in Fig. \ref{f.clean}.

 At low field, a particular feature of the spin model,
not discussed previously in the bosonic mapping, shows up.
Doping the dimer-singlet phase, we get long-range
antiferromagnetic ordering of the free moments appearing 
around each of the vacancies, a well-known phenomenon of 
\emph{order-by-disorder} \cite{ShenderK91,RoscildeH05}. 
This corresponds approximately to bosons
being created around $c$-sites but largely fluctuating
in number given that the chemical potential $\mu_c$ is 
very small, and coherently propagating between $c$-sites
through effective long-range hoppings which decay exponentially
with the inter-site distance, 
so that tenuous long-range phase coherence is established. 
Increasing the field
leads to a strong increase in the $\mu_c$  chemical 
potential, which quenches particle number fluctuations
and it destroys the disorder-induced superfluid order.
The system enters then a gapless quantum-disordered
phase, in which the gapless nature, reflected in the
finite compressibility of the bosons (susceptibility
of the magnetic system), is due to the fact that
$c$-sites are not all completely filled. For a larger
ratio $J/J'$ than the one considered here we can access
the regime where $c$-sites are essentially all filled
and the system does not admit any further particle,
namely the compressibility vanishes 
and the system acquires a gap as in the clean 
case \cite{RoscildeHprep}.
In this regime, either gapful or gapless, 
bosons are mostly pinned to $c$-sites,
and they cannot propagate in $b$-site regions due
to the local finite particle gap in such regions.

\begin{figure}[h!]
\begin{center}
\includegraphics[
     width=160mm,angle=0]{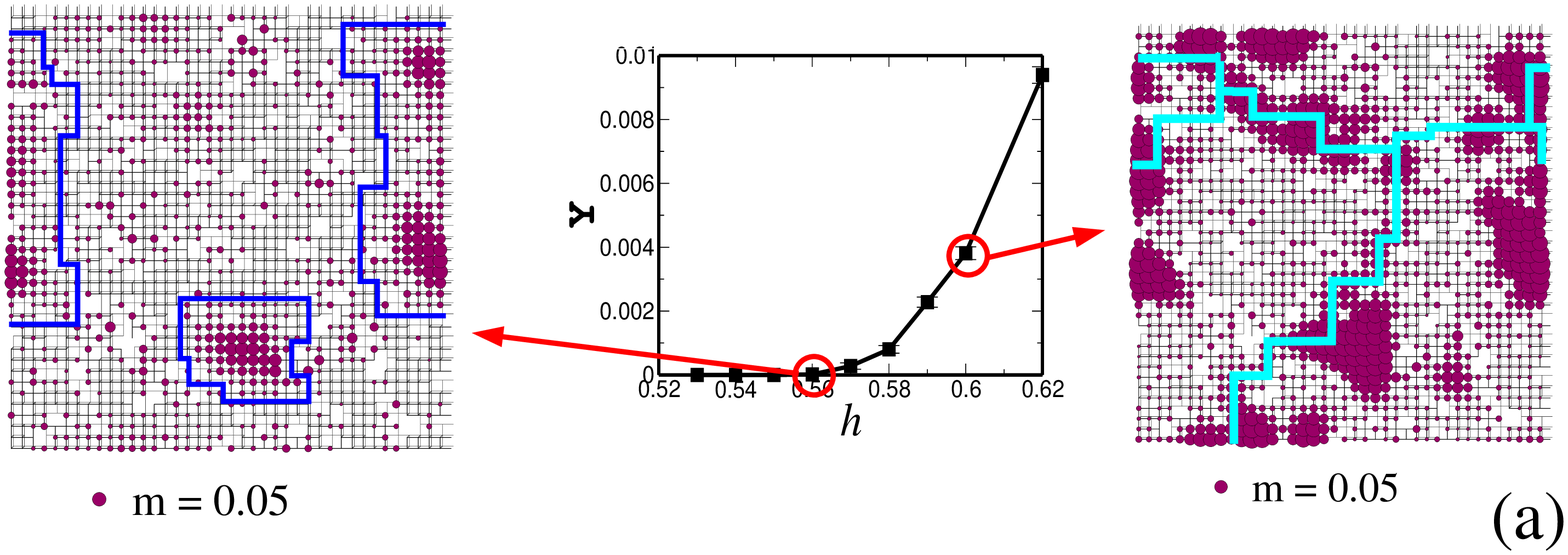} \\
 \includegraphics[
     width=160mm,angle=0]{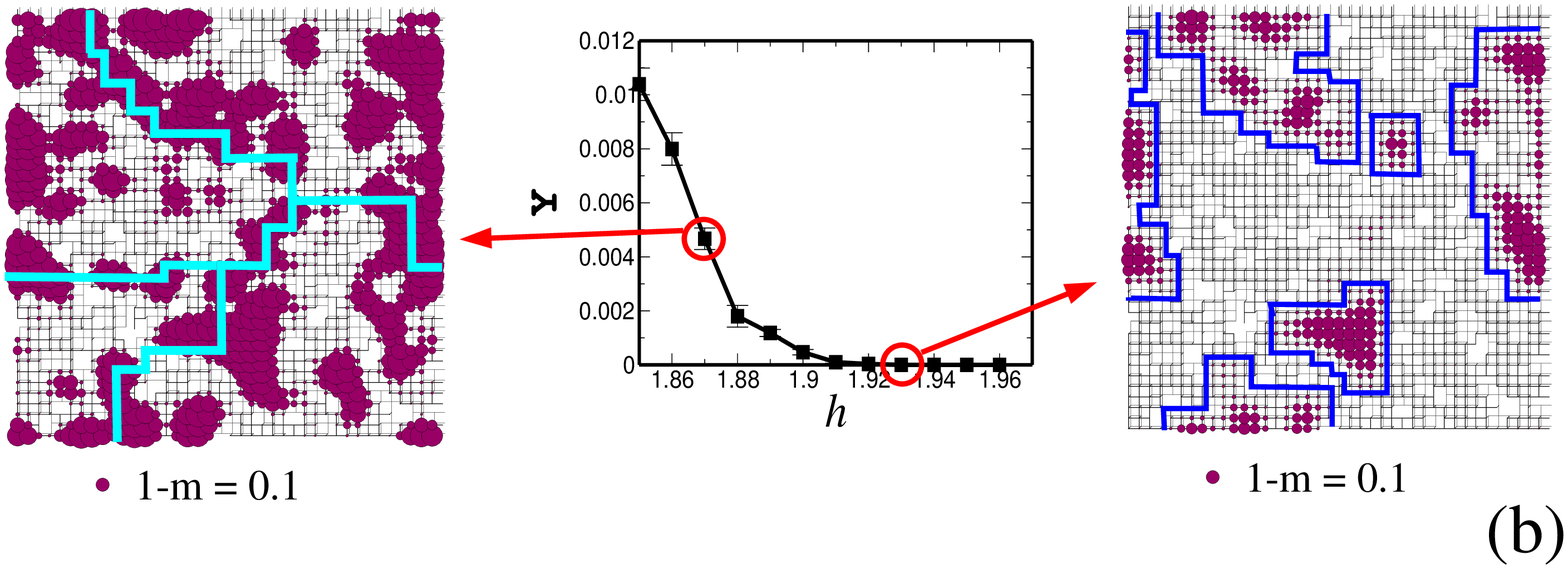}     
\caption{(a) Real-space images of the dimer
magnetization $m_i=\langle S^{z}_{i,1} + S^{z}_{i,2} \rangle$ 
on intact dimers in a 40x40x2 bilayer with $J/J'=4$, 
dilution $p=0.1$ and at
inverse temperature $\beta J = 256$, for $h = 0.56$
(\emph{left}) and $h = 0.6$ (\emph{right}). 
The radius of the dots is proportional
to the dimer magnetization.
The magnetization of unpaired spins ($c$-sites) is omitted for 
clarity. The most visible localized states are 
highlighted in the left panel, while the backbone of the percolating
magnetized network is highlighted in the right one.
The central panel shows the superfluid density 
as a function of the field for the specific sample considered.  
(b) Real-space images of the distance of the
dimer magnetization from its saturation value, $1-m_i$,
for $h=1.87$ (\emph{left}) and $h=1.93$ (\emph{right}).
Other symbols, parameters and explanations as in (a).}
\label{f.realspace}
\end{center}
\end{figure}

 When $h > h^{(0)}_{c1} \approx 0.47(1)$, nonetheless, large clean
 $b$-regions accept the appearence of bosons, which remain
 quantum-localized, giving rise to 
 a Bose-glass phase. This is clearly seen in real-space
 picture of the dimer magnetization 
 $m_i=\langle S^{z}_{i,1} + S^{z}_{i,2} \rangle$ in 
 Fig. \ref{f.realspace}(a) (left panel), representing the local 
 density of bosons on $b$-sites within the approximate bosonic
 mapping. We observe that, in this regime, 
 the order parameter $m_s^{x}$ and the superfluid 
 density are zero, and the number of bosons, corresponding
 to the uniform magnetization, increases extremely
 slowly with increasing field, corresponding to the gradual 
 appearence of particles in the rare clean regions of the system. 
 
 Increasing the field beyond $h_{c1} \approx 0.69(2) $ 
 a superfluid state is finally established in the system
 through delocalization of the collective state
 of the bosons [Fig. \ref{f.realspace}(a) (right panel)].
 Switching then to a description in terms of bosonic holes
 when the filling exceeds 1/2, we observe that the 
 delocalized superfluid state of holes, whose local
 concentration is proportional to $1-m_i$
 [Fig. \ref{f.realspace}(b) (left panel)], is destroyed
 at a critical field $h_{c2}\approx 1.78(2)$ lower than 
 the saturation one through a mechanism of quantum
 localization [Fig. \ref{f.realspace}(b) (right panel)],
 which gives rise to the hole Bose-glass phase at high
 fields. Even more clearly than in the lower-field case,
 this phase shows a finite compressibility and hence
 a gapless spectrum, but absence of long-range phase
 coherence due to quantum localization.

\section{Conclusions}

 In this work we have investigated the emergence of 
 a rich physical scenario of strongly correlated bosons 
 in weakly-coupled dimer antiferromagnets in a magnetic
 field. In presence of site dilution of the magnetic
 lattice, extended Bose-glass phases appear
 around the field-induced ordered phase, suggesting
 the possibility of an experimental realization 
 of such a phase in quantum magnets. 
  The particular case of a bilayer antiferromagnet 
  we considered in this paper
  is realized by BaCuSi$_2$O$_6$ \cite{Jaimeetal04}. 
  Nonetheless the picture of quantum localization 
  of bosonic quasiparticles applies straighforwardly to other  
  spin gap systems with different geometries, such as 
  Sr$_2$Cu(BO$_3$)$_2$ \cite{Sebastianetal05} or Tl(K)CuCl$_3$
  \cite{Rueggetal03}, where site dilution can be 
  analogously realized by doping the Cu sites. 
  The availability of high magnetic fields allows to scan
  the whole succession of phases discussed in the paper, 
  and different probes, such as elastic neutron scattering 
  and magnetometry measurements, can directly access the magnetic 
  observables characterizing the various phases. An open
  question is how to directly probe the superfluid 
  nature of the bosonic quasiparticles condensed in the
  ordered-ground state, given that a supercurrent of
  such particles corresponds to a pure \emph{spin} current with
  no energy or charge transport associated with it.

  \section{Acknowledgements}
  We thank M. Vojta for insightful discussions, and all the organizers
  of the Cortona BEC workshop 2005 for the exciting program they put together.
  This work is supported by DOE under grant DE-FG02-05ER46240. Computational 
  facilities have been generously provided by the HPC Center at USC.


\newpage
 
\begin{thebibliography}{99}
\bibitem{FazioZ01} R. Fazio and H. van der Zant, 
Phys. Rep.{\bf 355}, 235 (2001).
\bibitem{Greineretal02} M. Greiner,  
O. Mandel, T. Esslinger, T. W. H\"ansch, and I. Bloch, 
Nature (London) {\bf 415}, 39 (2002). 
\bibitem{Nikunietal00} T. Nikuni, M. Oshikawa, A. Oosawa, and H. Tanaka, 
 Phys. Rev. Lett. {\bf 84}, 5868 (2000).
\bibitem{Rueggetal03} Ch. R\"uegg N. Cavadini, A. Furrer, H.-U. 
G\"udel, K. Krämer, H. Mutka, A. Wildes, K. Habicht, and P. Vorderwisch,
Nature {\bf 423}, 62 (2003) , Nature (London) 
{\bf 423}, 62 (2003). 
\bibitem{Rice02} T.M. Rice, Science {\bf 298}, 760 (2002).  
\bibitem{Matsumotoetal04} 
M. Matsumoto, B. Normand, T. M. Rice, and M. Sigrist,
Phys. Rev. Lett. {\bf 89}, 077203 (2002);
Phys. Rev. B {\bf 69}, 054423 (2004).     
\bibitem{Fisheretal89} M.P.A. Fisher,
P. B. Weichman, G. Grinstein, and D. S. Fisher.   
Phys. Rev. B {\bf 40}, 546 (1989).
\bibitem{Scalettaretal91} R. T. Scalettar, G. G. Batrouni, 
and G. T. Zimanyi, Phys. Rev. Lett. {\bf 66}, 3144 (1991).
\bibitem{Vajketal02} O. P. Vajk, P. K. Mang, M. Greven, P. M. Gehring, 
and J. W. Lynn,
Science {\bf 295}, 1691 (2002).
\bibitem{Villain74} J. Villain, J. Phys. (Paris) {\bf 35}, 27 (1974).
\bibitem{Leggett95} A. J. Leggett, in \emph{Bose-Einstein Condensation}, 
ed. by A. Griffin, D. W. Snoke and S. Stringari (Cambridge, NY, 1995),
pag. 452.
\bibitem{SandvikS94} A. W. Sandvik and D. J. Scalapino, 
Phys. Rev. Lett. {\bf 72}, 2777 (1994).
\bibitem{Sommeretal01} T. Sommer, M. Vojta, and K. W. Becker, 
Eur. Phys. J. B {\bf 23}, 329 (2001).
\bibitem{Jaimeetal04} M. Jaime, V. F. Correa, 
N. Harrison, C. D. Batista, N. Kawashima, Y. Kazuma, G. A. Jorge,
R. Stern, I. Heinmaa, S. A. Zvyagin, Y. Sasago, and K. Uchinokura, 
Phys. Rev. Lett. {\bf 93}, 087203 (2004);
S. E. Sebastian, P. A. Sharma, M. Jaime, N. Harrison, V. Correa, 
L. Balicas, N. Kawashima, C. D. Batista, and I. R. Fisher,
Phys. Rev. B 72, 100404 (2005).  
\bibitem{TachikiY70} M. Tachiki and T. Yamada, 
J. Phys. Soc. Jpn. {\bf 28}, 1413 (1970).
\bibitem{SyljuasenS02} O.F. Sylju\aa sen and A.W. Sandvik,
Phys. Rev. E {\bf 66}, 046701 (2002).
\bibitem{PollockC87} E. L. Pollock and D. M. Ceperley,
Phys. Rev. B 36, 8343 (1987).
\bibitem{RoscildeH05} T. Roscilde and S. Haas,
Phys. Rev. Lett. {\bf 95}, 207206 (2005).   
\bibitem{Sandvik02}  A.W. Sandvik, Phys. Rev. B 
{\bf 66}, 024418 (2002).
\bibitem{ShenderK91} E.F. Shender and S.A. Kivelson, 
Phys. Rev. Lett. {\bf 66}, 2384 (1991).
\bibitem{RoscildeHprep} T. Roscilde and S. Haas,
in preparation.
\bibitem{Sebastianetal05} S.E. Sebastian, D. Yin, P. Tanedo, 
G. A. Jorge, N. Harrison, M. Jaime, Y. Mozharivskyj, G. Miller, 
J. Krzystek, S. A. Zvyagin, and I. R. Fisher, Phys. Rev. B 
{\bf 71}, 212405 (2005).

\end{thebibliography}
\end{document}